\documentclass[a4paper,12pt]{article}

\usepackage[english]{babel}
\usepackage[dvips]{graphicx}
\usepackage{verbatim}
\usepackage{makeidx}
\usepackage{color}
\usepackage{amsfonts}
\usepackage{amsmath}

\newcommand{\sqrtsNN}{\sqrt{s_{\scriptscriptstyle{{\rm NN}}}}}
\newcommand{\av}[1]{\left\langle #1 \right\rangle}

\newcommand{\mev}{\mathrm{MeV}}
\newcommand{\gev}{\mathrm{GeV}}
\newcommand{\tev}{\mathrm{TeV}}
\newcommand{\fm}{\mathrm{fm}}

\newcommand{\mum}{\mathrm{\mu m}}

\newcommand{\PbPb}{\mbox{Pb--Pb}}

\newcommand{\NN}{\mbox{nucleon--nucleon}}
\newcommand{\pp}{\mbox{proton--proton}}

\newcommand{\RAA}{R_{\rm AA}}
\newcommand{\RDh}{R_{{\rm D}/h}}
\newcommand{\pt}{p_{\rm t}}

\newcommand{\ccbar}{\mbox{$\mathrm {c\overline{c}}$}}

\newcommand{\Dz}{\mbox{$\mathrm {D^0}$}}
\newcommand{\DtoKpi}{\mbox{${\rm D^0\to K^-\pi^+}$}}

\newcommand{\Jpsi} {\mbox{J\kern-0.05em /\kern-0.05em$\psi$}\xspace}

\title{Probing the QGP with charm at ALICE$-$LHC\footnote{
This talk was presented in the New Talents Session at the 
``$41^{\rm st}$
International School of Subnuclear Physics, 2003'' in Erice (Italy)
and selected for publication in the proceedings of the School.
The present paper is an extract from Ref.~\cite{paperQuench}, 
where more details on the subject can be found.}
}

\author{Andrea Dainese\\
        \small Universit\`a degli Studi di Padova, 
        via Marzolo 8, 35131 Padova, Italy\\
        \small e-mail: andrea.dainese@pd.infn.it }
\date{}

\begin{document}

\maketitle

\begin{abstract}
\noindent The exclusive reconstruction of $\Dz$ mesons in the ALICE 
experiment allows to study the QCD energy loss of charm quarks in the 
deconfined quark--gluon plasma (QGP) medium expected to be produced in 
central nucleus--nucleus collisions at the Large Hadron Collider.



\end{abstract}

\section{Introduction}
\label{intro}

The ALICE experiment~\cite{tpalice} at the LHC 
will study nucleus--nucleus (AA) collisions at a centre-of-mass energy
$\sqrtsNN=5.5~\tev$ (for \PbPb) per nucleon--nucleon (NN) pair in order to 
investigate the properties of QCD matter at energy densities of 
few hundred times the density of atomic nuclei. In these conditions
a deconfined state of quarks and gluons is expected to be formed~\cite{karsch}.

Hard partons and heavy quarks, abundantly produced at LHC energies in 
initial hard scattering processes, are sensitive probes of the medium 
formed in the collision as they may lose energy by gluon bremsstrahlung 
while propagating through the medium itself. 
The attenuation (quenching) of leading hadrons and jets observed at 
RHIC~\cite{ludlam} is thought to be due to such a mechanism.
The large masses of the charm and beauty quarks make them qualitatively 
different probes, since, on well-established QCD grounds, in-medium energy loss
off massive partons is expected to be significantly smaller than off `massless'
partons (light quarks and gluons). 
Therefore, a comparative study of the attenuation of massless and 
massive probes is a promising tool to test the coherence of the interpretation 
of quenching effects as energy loss in a deconfined medium and to further 
investigate the properties (density) of such medium.

In the first part of this paper, we shortly summarize one of the widely 
used models of parton energy loss and we discuss how we used it in our 
simulation. In the second part, 
we show that the exclusive reconstruction of $\DtoKpi$ decays with ALICE 
allows to carry out the mentioned comparative quenching studies by
measuring the {\sl nuclear modification factor} of the D mesons 
transverse momentum ($\pt$) distribution
\begin{equation}
\label{eq:raa}
  R_{\rm AA}(\pt)\equiv
    \frac{{\rm d}N_{\rm AA}/{\rm d}\pt/{\rm binary~NN~collision}}
       {{\rm d}N_{\rm pp}/{\rm d}\pt},
\end{equation}
which would be 1 if the AA collision was a mere superposition of independent 
NN collisions without nuclear or medium effects, and the 
D/$charged~hadrons$ (D/$h$) ratio
\begin{equation}
\label{eq:RDh}
   R_{{\rm D}/h}(\pt)\equiv R_{\rm AA}^{\rm D}(\pt)\Big/R_{\rm AA}^h(\pt).
\end{equation}

\section{Parton energy loss and the dead cone effect for heavy quarks}
\label{energyloss}

In this work we use the 
quenching probabilities (or weights) calculated by C.A.~Salgado and 
U.A.~Wiedemann~\cite{carlosurs} in the framework of the `BDMPS' 
formalism~\cite{bdmps}, which we summarize in the following.
The energy loss obtained with the quenching weights is presented in 
Section~\ref{simulation}.

\begin{figure}[!t]
  \begin{center}
    \includegraphics[width=.49\textwidth]{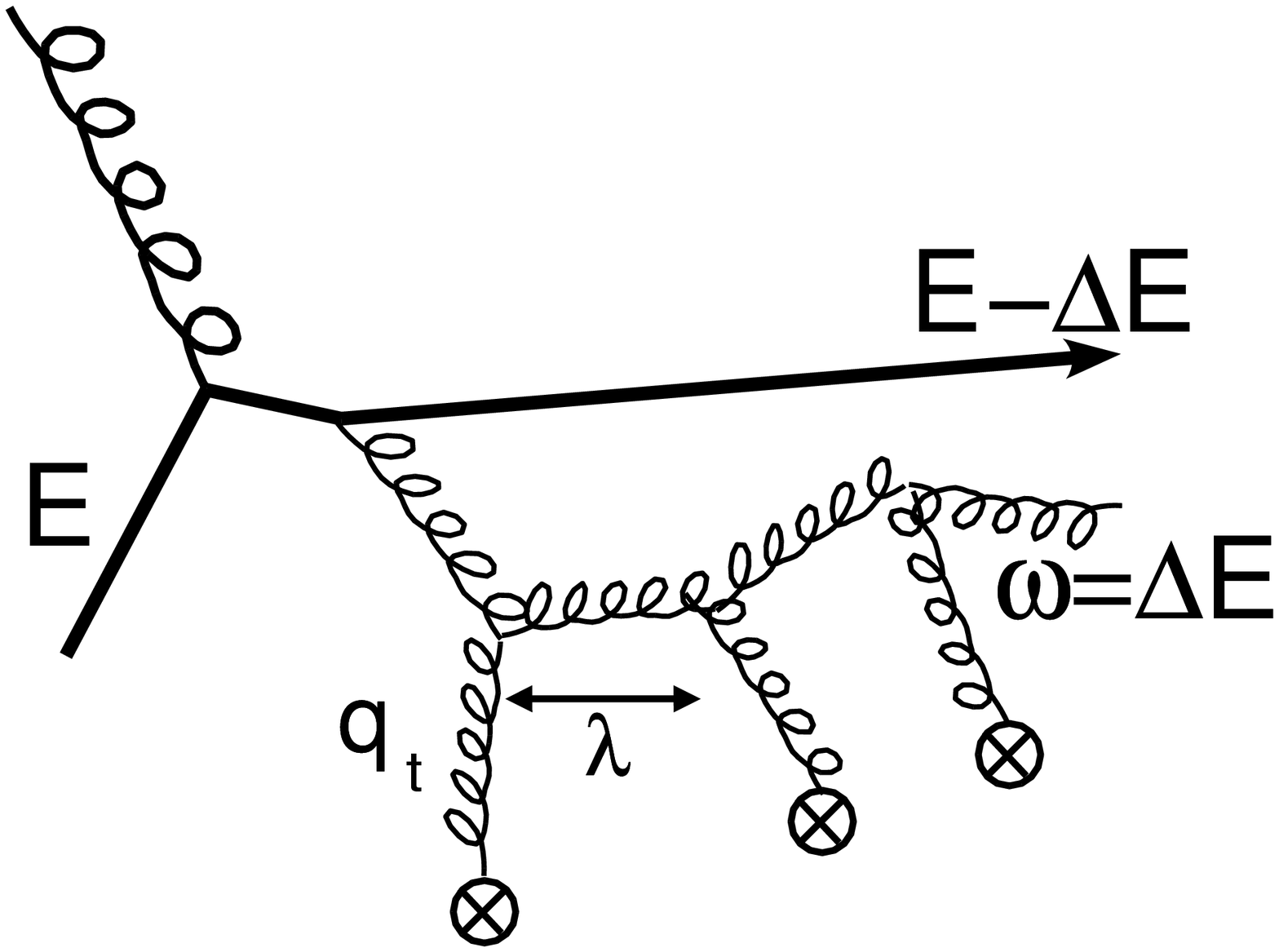}
    \includegraphics[width=.49\textwidth]{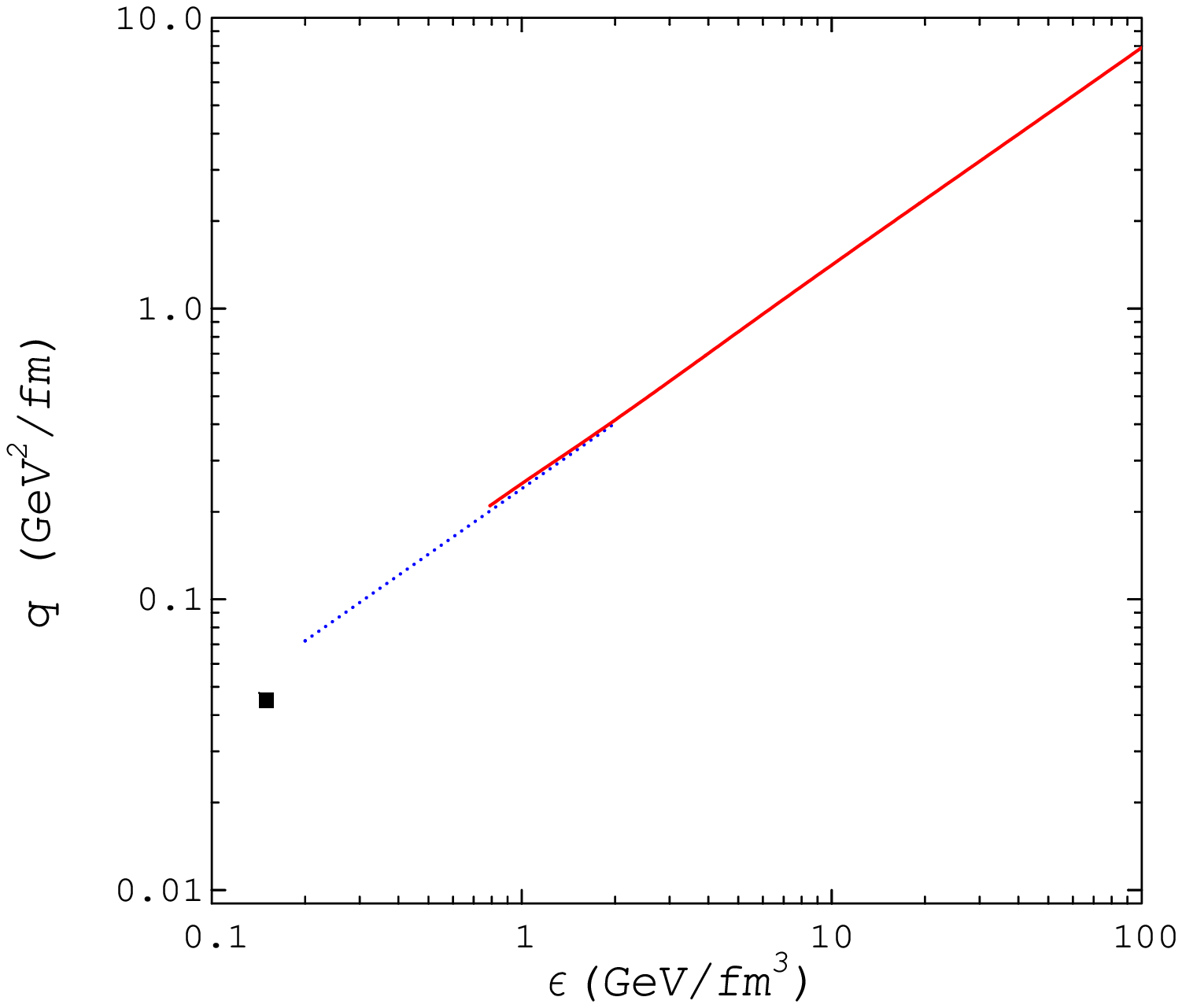}
    \caption{Typical gluon radiation diagram (left) and
             transport coefficient as a function of energy density (right) for 
             different media: cold (marker), massless hot pion gas 
             (dotted curve) and ideal QGP (solid curve)~\cite{baier}.}
    \label{fig:qtransp}
  \end{center}
\end{figure}

An energetic parton produced in a hard collision radiates a
gluon with a probability proportional to its path length $L$
in the dense medium. Then (Fig.~\ref{fig:qtransp}, left)
the radiated gluon undergoes multiple scatterings in the medium, in 
a Brownian-like motion with mean free path $\lambda$ which decreases 
as the density of the medium increases. The number of scatterings of the 
radiated gluon is also proportional to $L$. Therefore, the average energy loss
of the parton is proportional to $L^2$.

The scale of the energy loss is set by the `maximum' energy of the radiated
gluons, which depends on $L$ and on the properties of the 
medium:
\begin{equation}
 \omega_c = \hat{q}\,L^2/2,
\end{equation}
where $\hat{q}$ is the {\sl transport coefficient of the medium}, defined
as the average transverse momentum squared transferred to the projectile 
per unit path length, 
$\hat{q} = \av{q_{\rm t}^2}_{\rm medium}\big/\lambda$~\cite{carlosurs}.

In the case of a static medium, the distribution of the energy $\omega$ of the
radiated gluons (for $\omega\ll\omega_c$) is of the form:
\begin{equation} 
\label{eq:wdIdw}
\omega\frac{{\rm d}I}{{\rm d}\omega}\simeq \frac{2\,\alpha_s\,C_R}{\pi}\sqrt{\frac{\omega_c}{2\omega}},
\end{equation}
where $C_R$ is the QCD coupling factor (Casimir factor), equal 
to 4/3 for quark--gluon coupling and to 3 for gluon--gluon coupling.
The integral of the energy distribution up to $\omega_c$ estimates the 
average energy loss of the initial parton:
\begin{equation}
\label{eq:avdE}
\av{\Delta E} = \int^{\omega_c} \omega \frac{{\rm d}I}{{\rm d}\omega}{\rm d}\omega
\propto \alpha_s\,C_R\,\omega_c \propto \alpha_s\,C_R\,\hat{q}\,L^2.
\end{equation}
The average energy loss is therefore:
proportional to $\alpha_s\,C_R$ and, thus, larger by a factor 
$9/4=2.25$ for gluons than for quarks; 
proportional to the transport coefficient of the medium;
proportional to $L^2$; 
independent of the parton initial energy.  
The last point is peculiar to the BDMPS model. Other 
models~\cite{gyulassywang,gyulassy2} consider 
an explicit dependence of $\Delta E$ on the initial energy $E$. However, 
there is always an intrinsic 
dependence of the radiated energy on the initial energy, determined by the 
fact that the former cannot be larger than the latter, $\Delta E\leq E$.

The transport coefficient is proportional to the density of the 
scattering centres and to the typical momentum transfer in the 
gluon scattering off these centres.
Figure~\ref{fig:qtransp} (right) 
reports the estimated dependence of $\hat{q}$ on 
the energy density $\epsilon$ for different equilibrated 
media~\cite{baier}: for cold nuclear matter (marker) the estimate is 
$\hat{q}_{\rm cold} \simeq 0.05~\gev^2/\fm$; for a QGP formed at the LHC with 
$\epsilon\sim 50$--$100~\gev/\fm^3$, $\hat{q}$ is expected to be 
of $\simeq 5$--$10~\gev^2/\fm$. 

In Ref.~\cite{dokshitzerkharzeev} Yu.L.~Dokshitzer and D.E.~Kharzeev 
argue that for heavy quarks, 
because of their large mass, the radiative energy loss should be 
lower than for light quarks. The predicted consequence of this effect is
an enhancement of the ratio of D mesons to pions (or hadrons in general) 
at moderately 
large ($5$--$10~\gev/c$) transverse momenta, with respect to what
observed in the absence of energy loss (\pp~collisions).

Heavy quarks with momenta up to $20$--$30~\gev/c$ propagate with 
a velocity which is smaller than the velocity of light. As a consequence, 
gluon radiation at angles $\Theta$ smaller than the ratio of their mass to 
their energy $\Theta_0=m/E$ is suppressed by destructive quantum 
interference. 
The relatively depopulated cone around the heavy quark direction with 
$\Theta<\Theta_0$ is indicated as `dead cone'~\cite{dokshitzerdeadcone}.

In Ref.~\cite{dokshitzerkharzeev} the dead cone effect is assumed to 
characterize also in-medium gluon radiation and 
the energy distribution of the radiated gluons (\ref{eq:wdIdw}),
for heavy quarks, is estimated to be suppressed by the factor:
\begin{equation}
  \label{eq:FHL}
  \frac{{\rm d}I}{{\rm d}\omega}\bigg|_{\rm Heavy}\bigg/
  \frac{{\rm d}I}{{\rm d}\omega}\bigg|_{\rm Light}=
  \left[1+\frac{\Theta_0^2}{\Theta^2}\right]^{-2}=
  \left[1+\left(\frac{m}{E}\right)^2\sqrt{\frac{\omega^3}{\hat{q}}}\right]^{-2}
  \equiv F_{\rm H/L},
\end{equation} 
where the expression for the characteristic gluon emission 
angle~\cite{dokshitzerkharzeev} 
$\Theta\simeq (\hat{q}/\omega^3)^{1/4}$ has been used.
The heavy-to-light suppression factor $F_{\rm H/L}$ in (\ref{eq:FHL}) 
increases (less suppression) as the 
heavy quark energy $E$ increases (the mass becomes negligible) 
and it decreases at large $\omega$, indicating that the high-energy 
part of the 
gluon radiation spectrum is drastically suppressed by the dead cone effect.

\section{Simulation of energy loss}
\label{simulation}

The Salgado--Wiedemann (SW) quenching weight is defined as the probability 
that a hard parton radiates an energy $\Delta E$ due to scattering in 
spatially extended QCD matter. In Ref.~\cite{carlosurs}, the weights are 
calculated on the basis of the BDMPS formalism, keeping into account both 
the finite in-medium path length $L$ and the dynamic expansion of 
the medium after the nucleus--nucleus collision.
The input parameters for the calculation are the 
length $L$, the transport coefficient $\hat{q}$ and the parton species
(light quark or gluon).

The distribution of the in-medium path length in the plane transverse 
to the beam line\footnote{Partons produced at central rapidities 
propagate in the transverse plane.} for central
\PbPb~collisions (impact parameter $b<3.5~\fm$, corresponding to the 5\% most 
central collisions) is calculated in the framework
of the Glauber model of the collision geometry~\cite{glauber}. 
For a given impact parameter,
hard parton production points are sampled according 
to the density $\rho_{\rm coll}(x,y)$ of binary nucleon--nucleon collisions in 
the transverse plane and their 
azimuthal propagation directions are sampled uniformly. 
For a parton with production point $(x_0,y_0)$ 
and azimuthal direction  $(u_x,u_y)$, 
the path length is defined as:
\begin{equation}
\label{eq:ell}
L = \frac{\int_0^\infty {\rm d}l\,l\,\rho_{\rm coll}(x_0+l\,u_x,y_0+l\,u_y)}
     {0.5\,\int_0^\infty {\rm d}l\,\rho_{\rm coll}(x_0+l\,u_x,y_0+l\,u_y)}.
\end{equation} 

\begin{figure}[!t]
  \begin{center}
    \includegraphics[width=.54\textwidth]{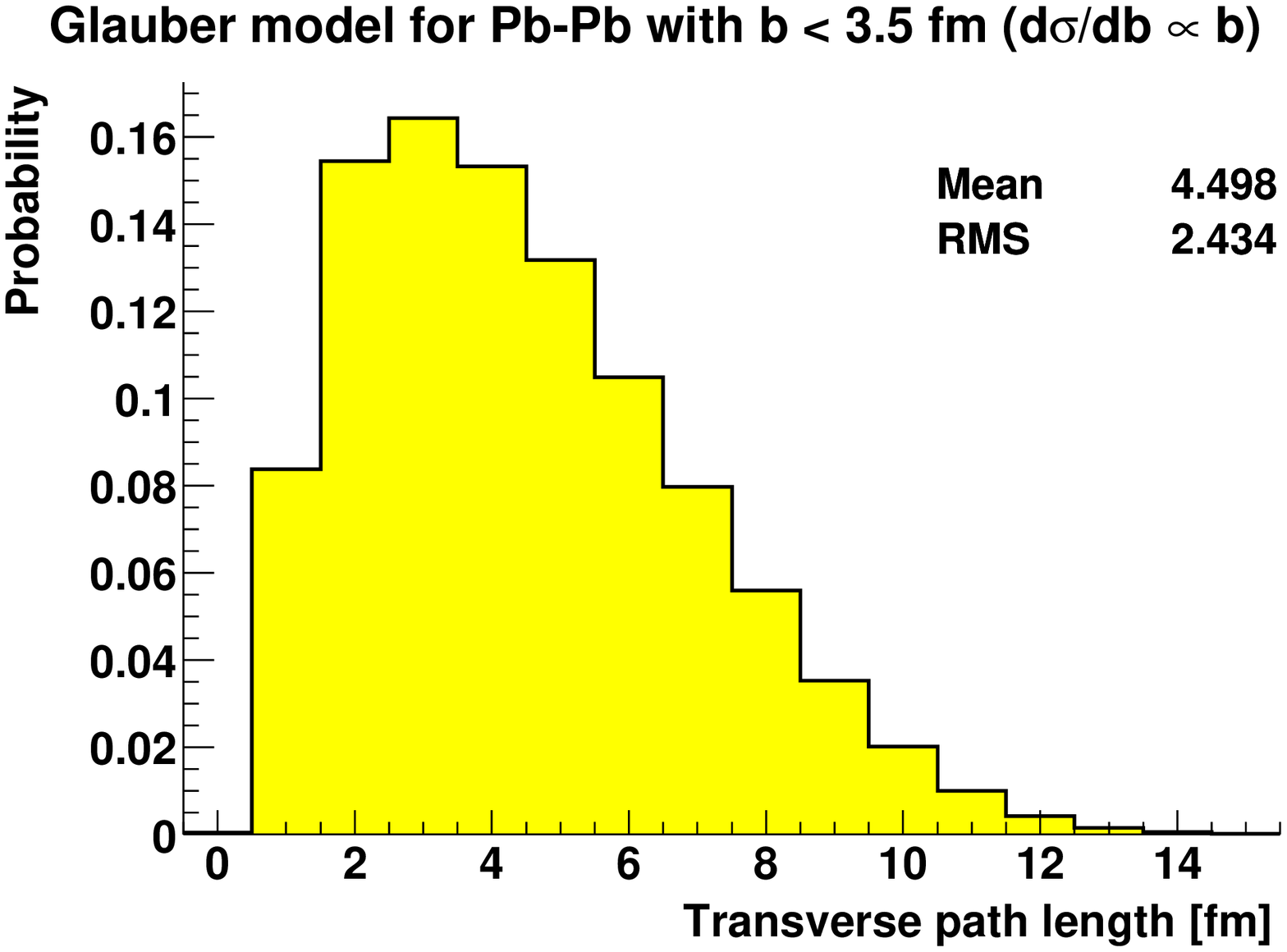}
    \includegraphics[width=.45\textwidth]{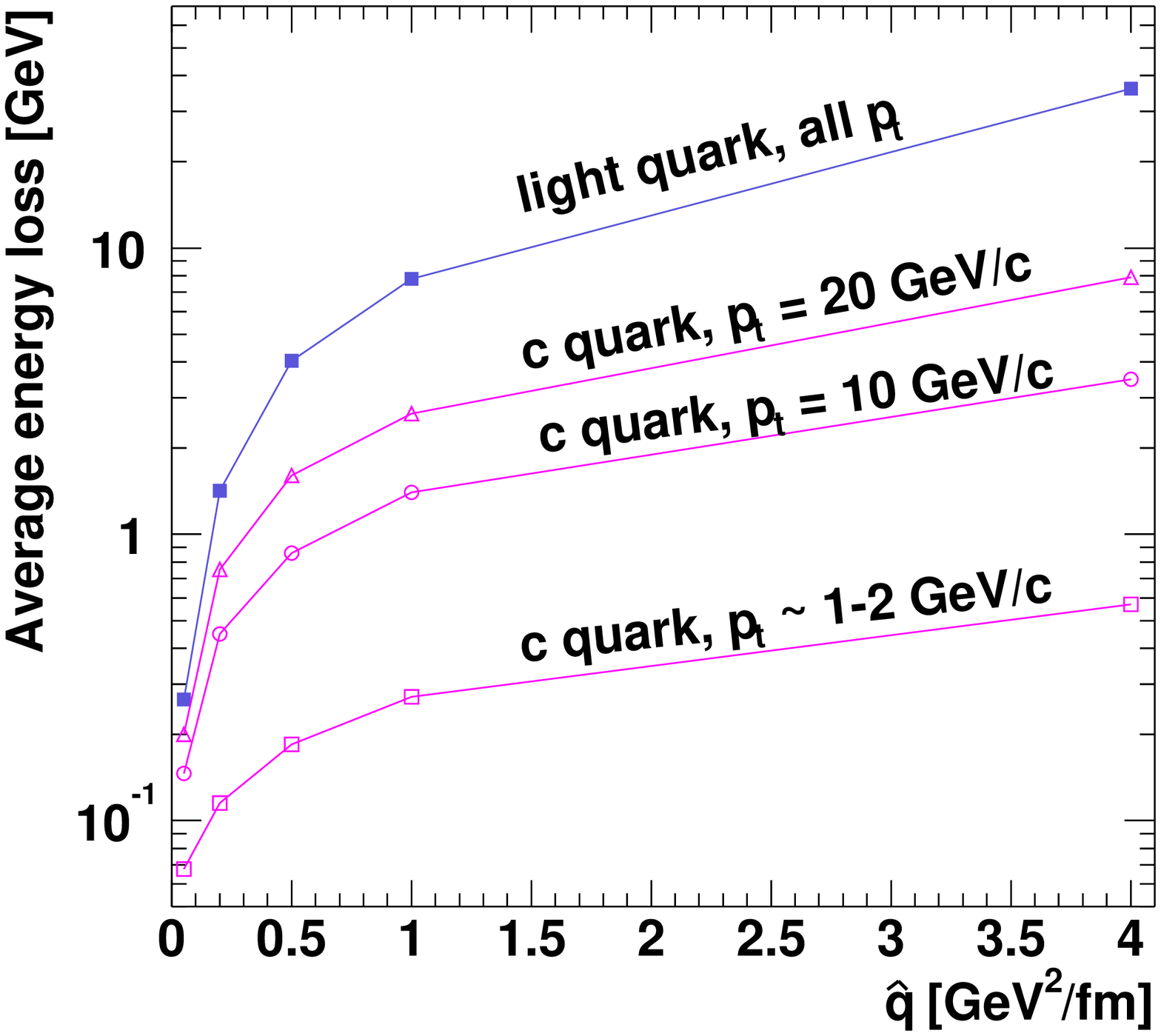}
    \caption{Distribution of the path lengths in the transverse plane 
             for partons produced in \PbPb~collisions with $b<3.5~\fm$ (left).
             Average energy loss as a function of the transport 
             coefficient (right).} 
    \label{fig:ellANDavdE}
  \end{center}
\end{figure}

Many sampling iterations are performed varying the impact 
parameter $b$ from $0.25~\fm$ to $3.25~\fm$ in steps of $0.5~\fm$. 
The obtained distributions are given a weight $b$, since we verified that 
d$\sigma^{\rm hard}/$d$b\propto b$ for $b<3.5~\fm$, and added together. 
The result is shown in Fig.~\ref{fig:ellANDavdE} (left). 
The average length is $4.5~\fm$, corresponding to 
about 70\% of the radius of a Pb nucleus and 
the distribution is significantly accumulated towards low values of $L$
because a large fraction of the partons are produced in the periphery
of the superposition region of the two nuclei (`corona' effect).

For a given value of the transport coefficient $\hat{q}$ and a given parton 
species, we use the routine 
provided in Ref.~\cite{carlosurs} to get the energy loss probability 
distribution $P(\Delta E;L)$ for the integer values of $L$ up to 
$15~\fm$. Then, these 15 distributions are 
weighted according to the path length probability in Fig.~\ref{fig:ellANDavdE}
and added together to obtain a global energy loss probability distribution 
$P(\Delta E)$. The energy loss to be used for the quenching simulation
can be directly sampled from the $P(\Delta E)$ distribution 
corresponding to the chosen $\hat{q}$ and to the correct parton species.

The predicted lower energy loss for charm quarks is accounted for  
by multiplying the $P(\Delta E)$ distribution for light 
quarks with the dead cone suppression factor $F_{\rm H/L}$ in (\ref{eq:FHL}).
It was verified that this approximation is equivalent to 
recalculating the SW quenching weights with the gluon energy distribution 
for heavy quarks as given in (\ref{eq:FHL})~\cite{carlosurspriv,thesis}.
Since $F_{\rm H/L}$ depends on the heavy quark energy, the
product has to be done for each c quark or, more conveniently, in bins of 
$\pt$.
Figure~\ref{fig:ellANDavdE} (right) reports the average energy 
loss as a function of the transport coefficient for light quarks and for 
charm quarks ($m_{\rm c}=1.2~\gev$) 
with $\pt= 1$--$2,~10,~20~\gev/c$, as obtained with the 
described dead cone correction ($\pt$-dependent
$P(\Delta E)\otimes F_{\rm H/L}$ product). With $\hat{q}=4~\gev^2/\fm$,
our estimated transport coefficient for the LHC (see next paragraph), 
the average energy loss for light quarks is $\av{\Delta E}\simeq 35~\gev$
(the effective value of $\av{\Delta E}$ is lower by about a factor 2, 
due to the constraint $\Delta E\leq E$).
For c quarks of $1$--$2,~10,~20~\gev/c$ $\av{\Delta E}$ is about 2\%, 10\% 
and 20\%, respectively, of the average loss for light quarks.

For the estimation of the transport coefficient $\hat{q}$ for our simulation, 
we consider that it is reasonable to require for central nucleus--nucleus 
collisions at the LHC a quenching of
hard partons at least of the same magnitude as that observed at RHIC.
We, therefore, derive the nuclear modification factor $\RAA$ for charged 
hadrons produced at the LHC and we choose the transport coefficient in 
order to obtain $\RAA\simeq 0.2$--$0.3$ in the range $\pt=5$--$10~\gev/c$
(for RHIC results see e.g. Refs.~\cite{ludlam,phenix}). 

The transverse momentum distributions, for $\pt>5~\gev/c$, of charged hadrons 
are generated by means of the chain:
\begin{enumerate}
\item generation of a parton, quark or gluon, with $\pt>5~\gev/c$, 
      using PYTHIA~\cite{pythia} proton--proton 
      with $\sqrt{s}=5.5~\tev$ and CTEQ 4L 
      parton distribution functions; with these parameters, the parton 
      composition given by PYTHIA is 78\% gluons and 22\% quarks;
\item sampling of an energy loss $\Delta E$ according to $P(\Delta E)$ 
      and calculation of the quenched 
      transverse momentum of the parton, $\pt'=\pt-\Delta E$ (if 
      $\Delta E>\pt$, $\pt'$ is set to 0);
\item (independent) fragmentation of the parton to a hadron using the 
      leading order Kniehl-Kramer-P\"otter (KKP) fragmentation 
      functions~\cite{kkp}. 
\end{enumerate}
Quenched and unquenched $\pt$ distributions are obtained including or 
excluding the second step of the chain. $\RAA$ is calculated as 
the ratio of the $\pt$ distribution with quenching to the $\pt$ distribution 
without quenching. We find $\RAA\simeq 0.25$--$0.3$ in $5<\pt<10~\gev/c$ for 
$\hat{q}=4~\gev^2/\fm$. This value is reasonable, as it corresponds, 
using the plot in 
Fig.~\ref{fig:qtransp}, to an energy density 
$\epsilon\simeq 40$--$50~\gev/\fm^3$, which is about a factor 2 lower 
than the maximum energy density expected for central \PbPb~collisions at the
LHC.

Charm quarks are generated using PYTHIA, tuned in order to reproduce the 
single-inclusive c (and $\overline{\rm c}$) $\pt$ distribution predicted 
by the pQCD program 
HVQMNR~\cite{hvqmnr} with $m_{\rm c}=1.2~\gev$ and 
$\mu_{\rm Fact.}=\mu_{\rm Renorm.}=2\,m_{\rm t}\equiv 2\,\sqrt{m_{\rm c}^2+\pt^2}$
(the details on this tuning can be found in Ref.~\cite{noteHVQprod}).
We use the CTEQ 4L parton distribution functions including the nuclear
shadowing effect by means of the EKS98 parameterization~\cite{EKS}
and the parton intrinsic transverse momentum broadening as reported
in Ref.~\cite{noteHVQprod}. 
Energy loss for charm quarks is simulated following a slightly different 
procedure with respect to that for light quarks and gluons. Since
the total number of $\ccbar$ pairs per event has to be conserved, in the 
cases where the sampled $\Delta E$ is larger than $\pt$, we assume the c 
quark to be thermalized in the medium and we give it a transverse momentum 
according to the distribution 
d$N/$d$m_{\rm t}\propto m_{\rm t}\,\exp(-m_{\rm t}/T)$. 
We use $T=300~\mev$ as the thermalization temperature. The other difference
with respect to the previous case is that we use the standard string 
model in PYTHIA for the c quark fragmentation.

\section{Charm reconstruction with ALICE}
\label{D0reco}

The transverse momentum distribution of charm mesons produced at central 
rapidity, $|y|<1$, can be directly measured with ALICE from the exclusive 
reconstruction of $\DtoKpi$ (and charge conjugates). 
The displaced vertices of $\Dz$
decays ($c\tau=124~\mum$) can be identified in the ALICE Inner Tracking 
System, that provides a measurement of the track impact parameters to 
the collision vertex with a resolution better than $50~\mum$ for 
$\pt>1~\gev/c$. The low value of the magnetic field (0.4~T) and the 
K/$\pi$ separation in the ALICE Time of Flight allow to extend 
the measurement of the $\Dz$ production cross section down to almost 
0 transverse momentum. The strategy for this analysis and the
selection cuts to be applied were studied with a realistic and detailed 
simulation of the detector geometry and response, including the main
background sources~\cite{thesis,D0jpg}. 

\begin{figure}[!t]
  \begin{center}
    \includegraphics[width=.49\textwidth]{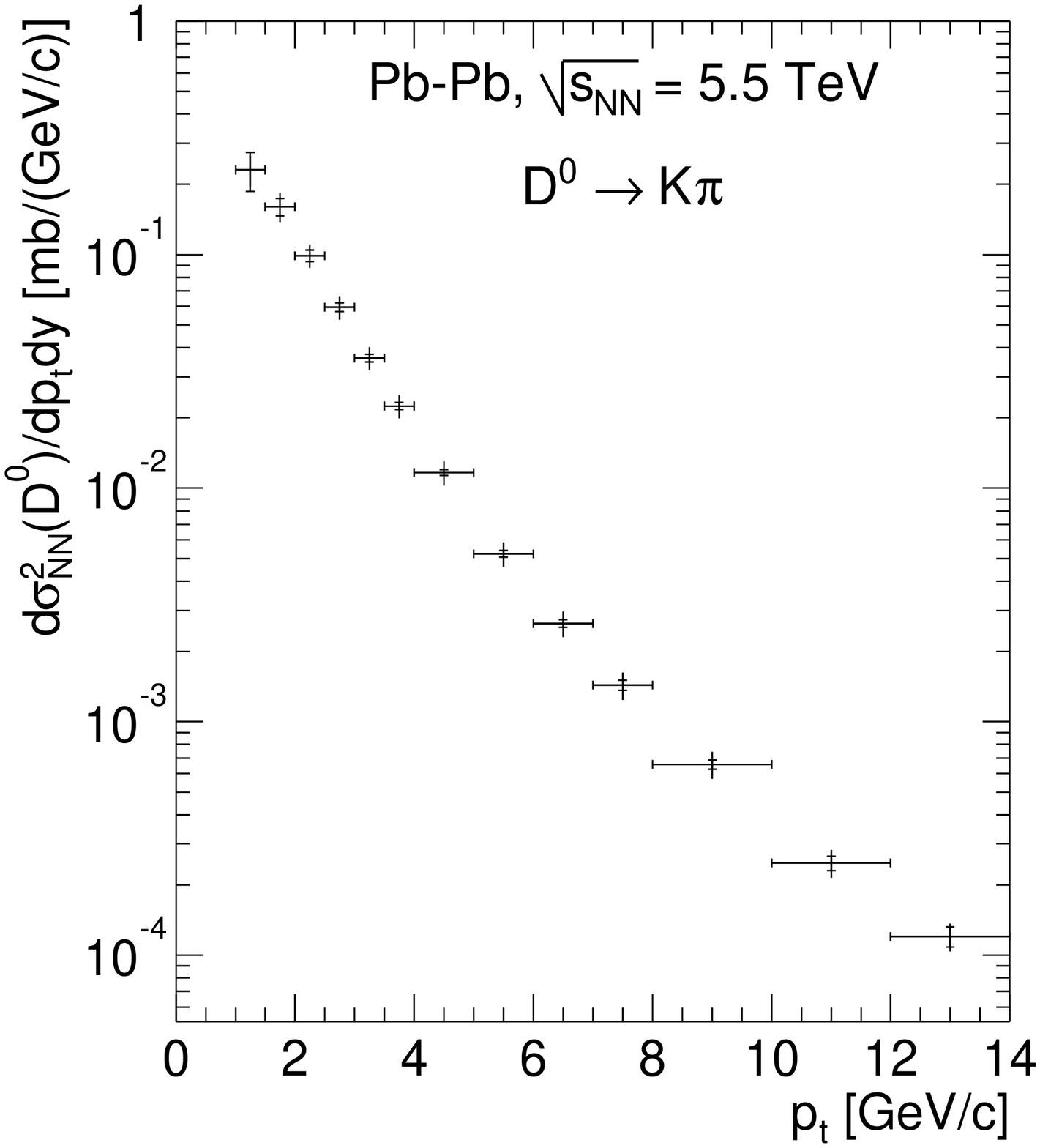}
    \includegraphics[width=.49\textwidth]{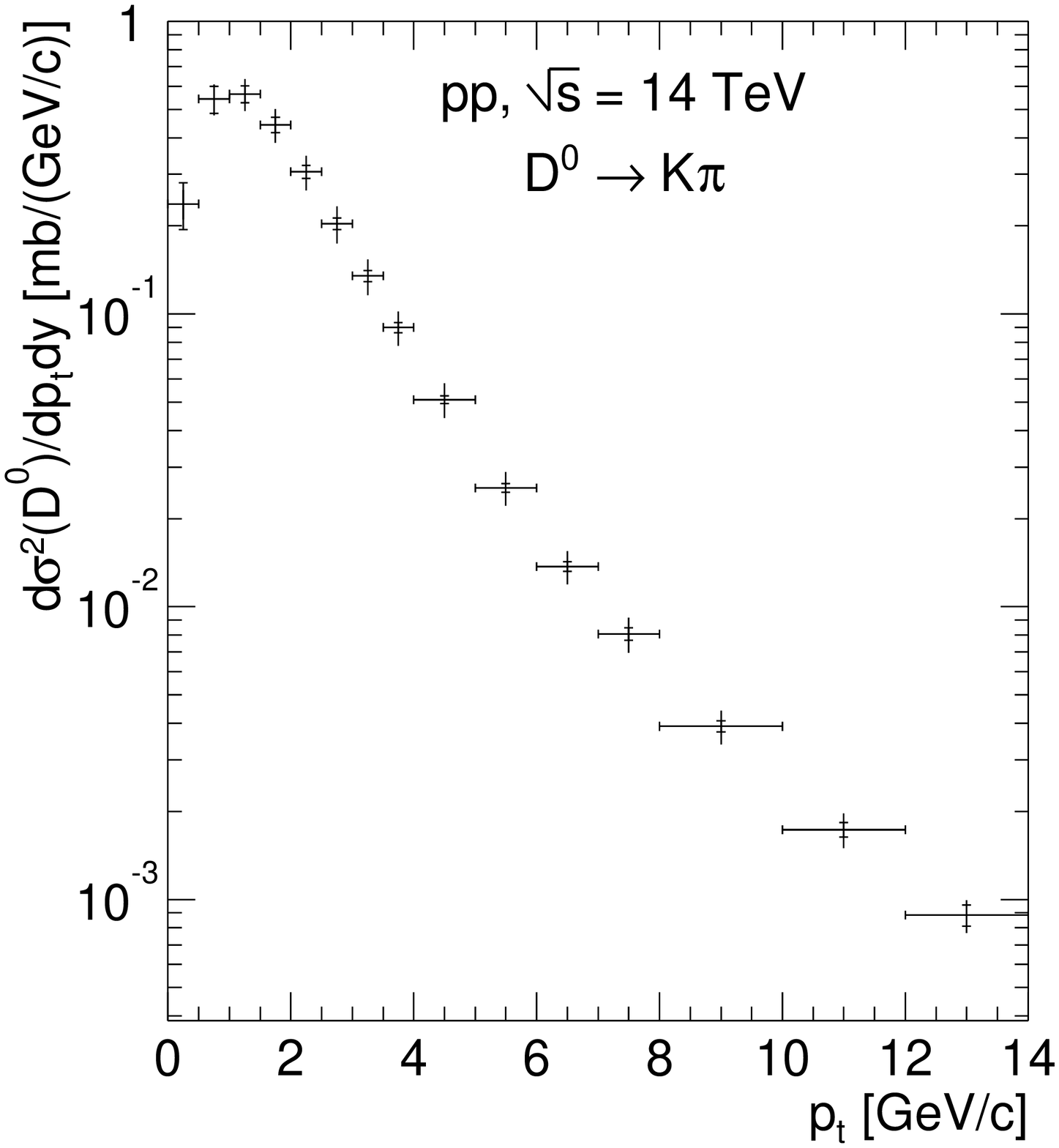}
    \caption{Double differential cross section per \NN~collision 
             for $\Dz$ production as a 
             function of $\pt$, as it can be measured with $10^7$ central  
             \PbPb~events (left) and $10^9$ pp minimum-bias events (right). 
             Statistical (inner bars) and $\pt$-dependent 
             systematic errors (outer bars) are shown. A normalization error
             of 11\% for \PbPb~and 5\% for pp is not shown.} 
    \label{fig:D0pt}
  \end{center}
\end{figure}

The expected performance for central \PbPb~($b<3.5~\fm$) 
at $\sqrtsNN=5.5~\tev$ and pp collisions at $\sqrt{s}=14~\tev$, 
as obtained using the input production yields 
$N^{\scriptstyle{\rm c\overline{c}}}_{\rm Pb-Pb}=115$ and
$N^{\scriptstyle{\rm c\overline{c}}}_{\rm pp}=0.16$ 
(see Ref.~\cite{noteHVQprod}), 
is summarized in Fig.~\ref{fig:D0pt}. The accessible 
$\pt$ range is $1$--$14~\gev/c$
for \PbPb~and $0.5$--$14~\gev/c$ for pp. In both cases the statistical error 
(corresponding to 1 month of data-taking for Pb--Pb and to 9 months for pp) 
is better than 15--20\% and the systematic error 
(acceptance and efficiency corrections, subtraction of the feed-down from 
${\rm B}\to \Dz+X$ decays, cross section normalization, 
centrality selection for \PbPb) is better than 20\%. More details 
are given in Ref.~\cite{thesis}.

\section{Results: $\RAA$ and $\RDh$}
\label{results}

\begin{figure}[!t]
  \begin{center}
    \includegraphics[width=0.49\textwidth]{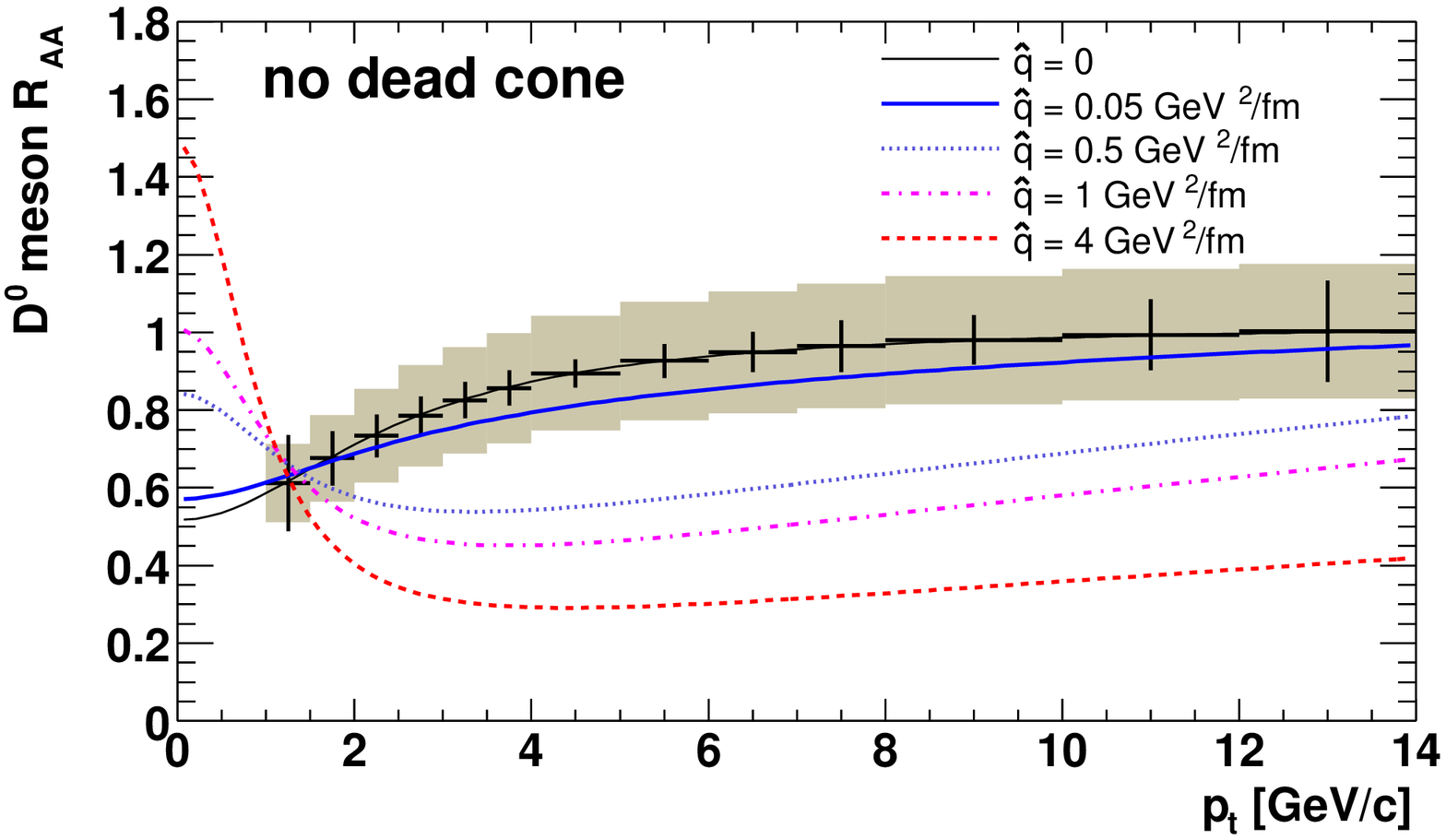}
    \includegraphics[width=0.49\textwidth]{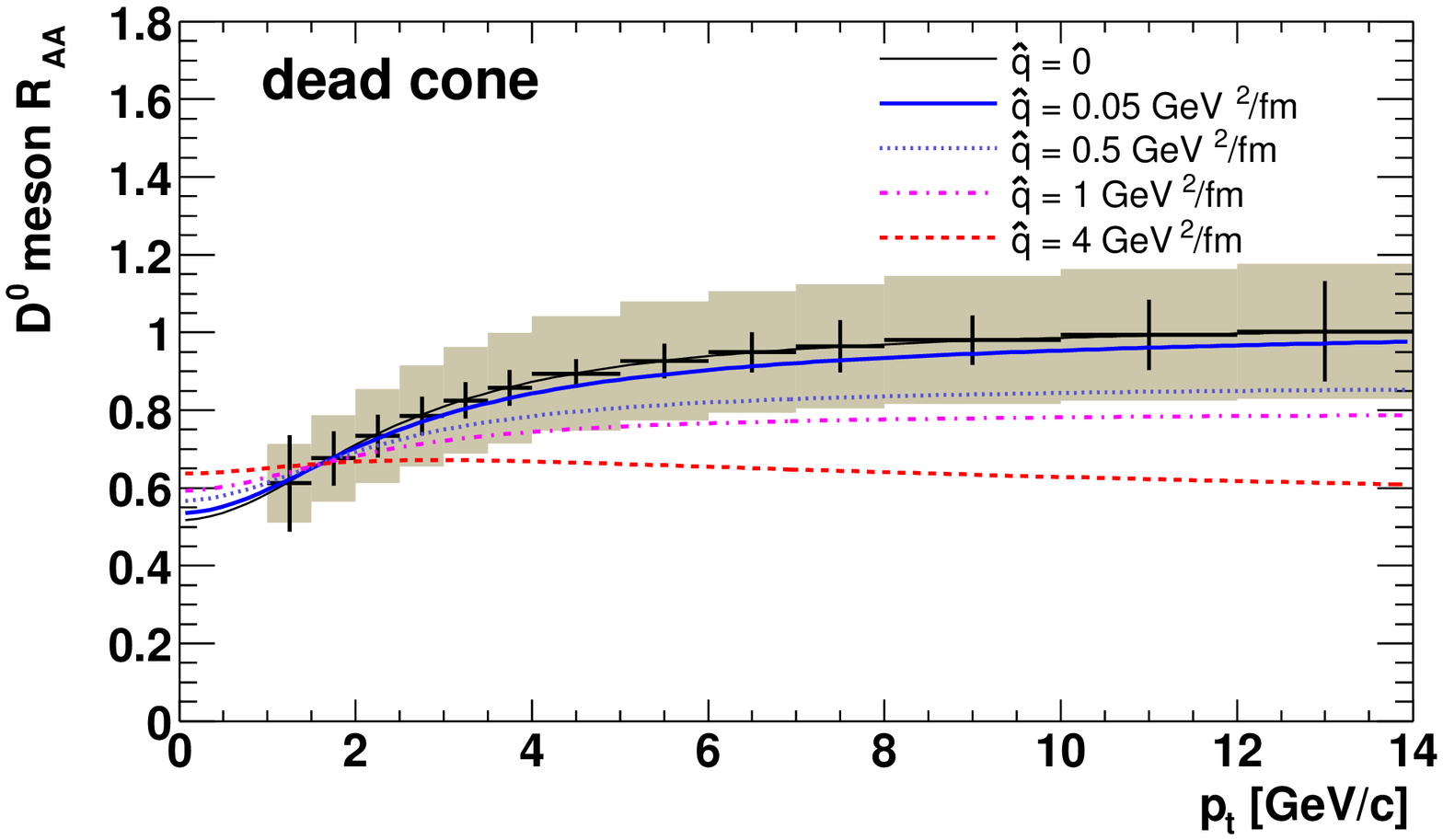}
    \caption{Nuclear modification factor for $\Dz$ mesons with shadowing, 
             intrinsic $k_{\rm t}$ broadening and parton energy loss.
             Left panel: without dead cone correction; right panel: with 
             dead cone correction. Errors corresponding to the curve 
             for $\hat{q}=0$ are shown: bars = statistical, 
             shaded area = systematic.} 
    \label{fig:RAA}
\vglue0.5cm
    \includegraphics[width=0.49\textwidth]{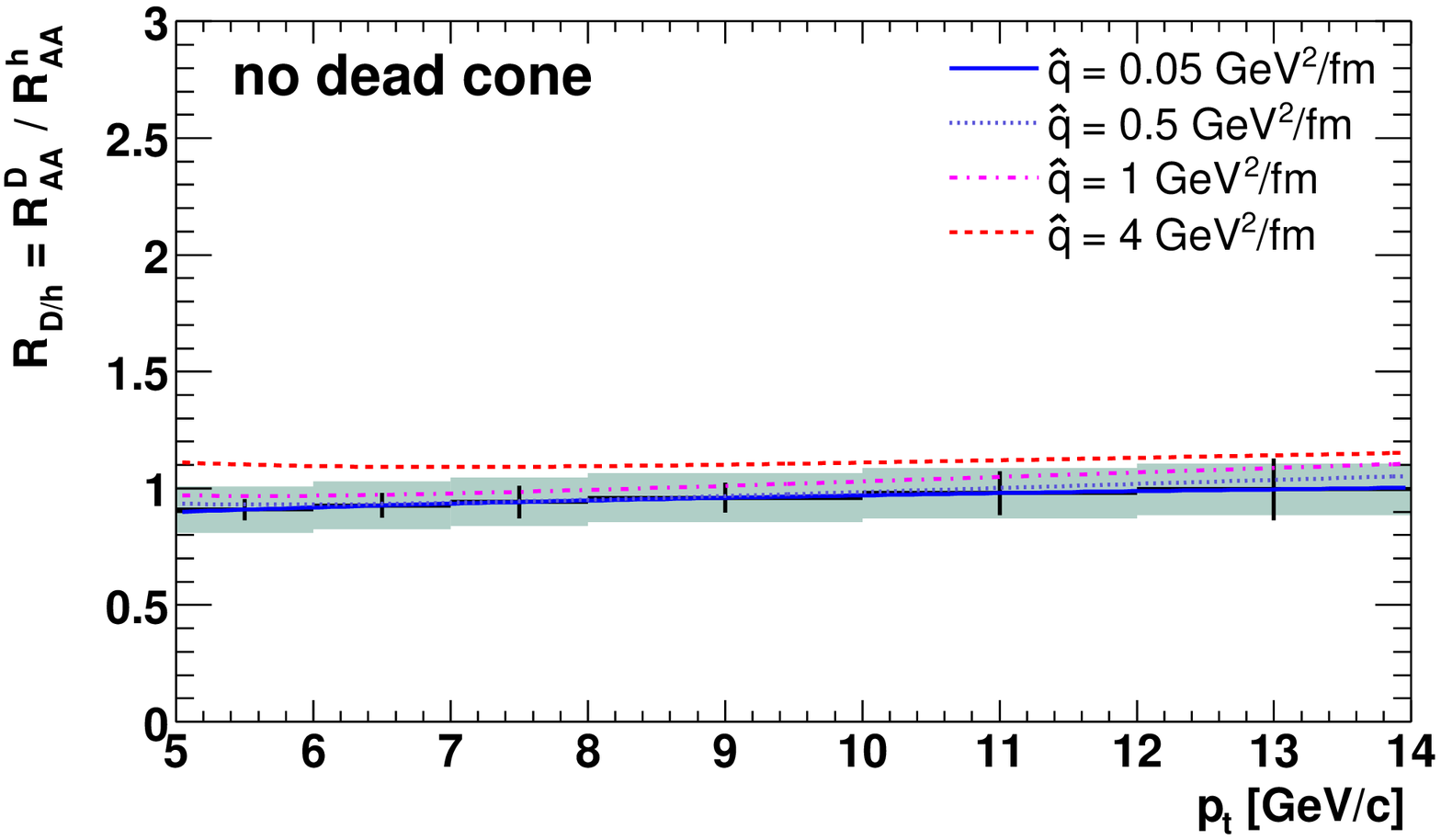}
    \includegraphics[width=0.49\textwidth]{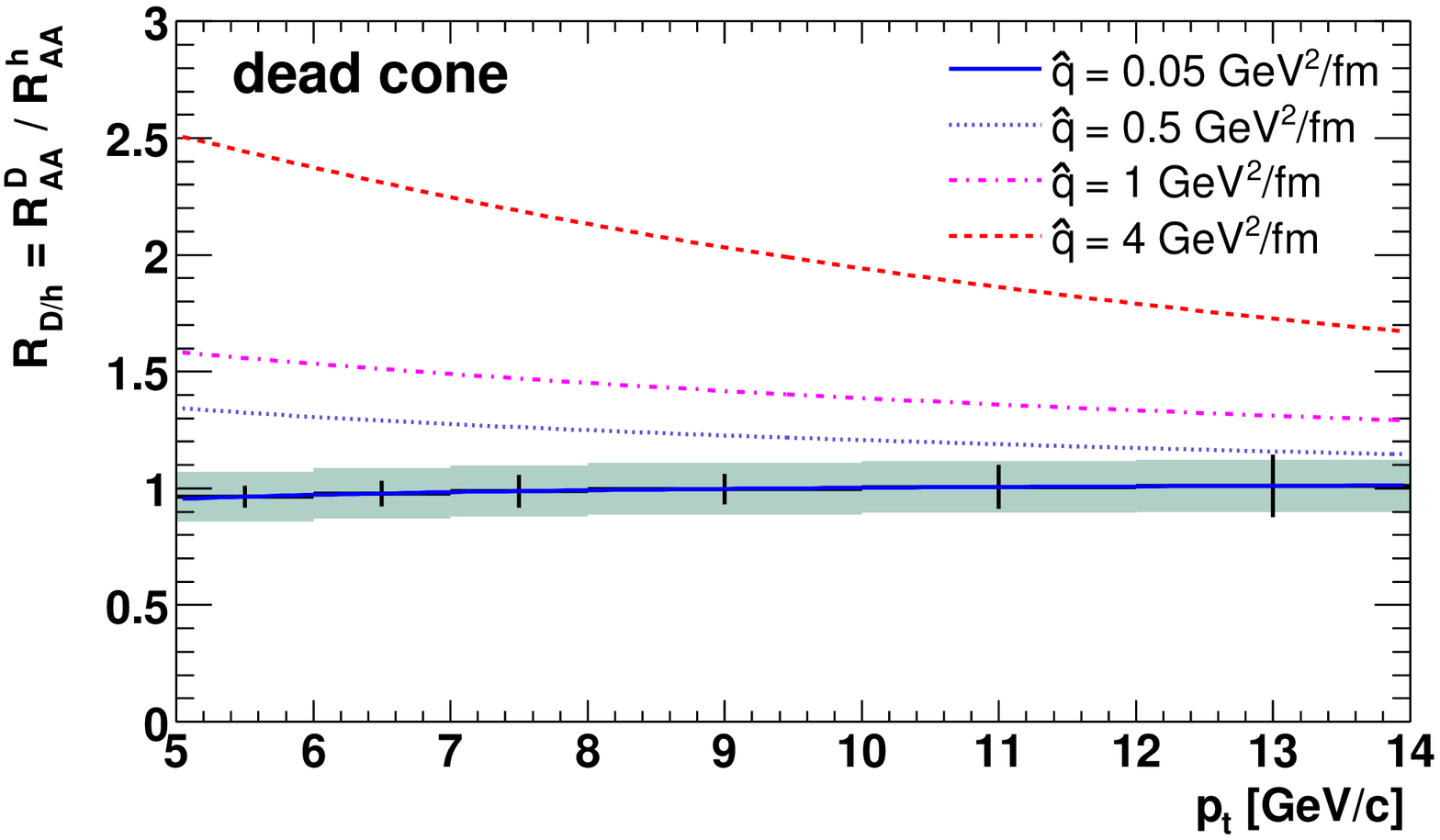}
    \caption{Ratio of the nuclear modification factors for $\Dz$ mesons 
             and for charged hadrons.
             Left panel: without dead cone correction; right panel: with 
             dead cone correction. Errors corresponding to the curve 
             for $\hat{q}=0.05~\gev^2/\fm$ are shown: bars = statistical, 
             shaded area = systematic.} 
    \label{fig:RDh}
  \end{center}
\end{figure}

The nuclear modification factor for $\Dz$ mesons is reported in 
Fig.~\ref{fig:RAA}. Nuclear shadowing, parton intrinsic 
transverse momentum broadening and energy loss are included.
The dead cone effect is not included in the left-hand panel and included in 
right-hand panel. Different values of the transport coefficient are used 
for illustration; we remind that the value expected on the basis of the pion 
quenching observed at RHIC is $\hat{q}=4~\gev^2/\fm$. The reported 
statistical (bars) and systematic (shaded area) errors are obtained combining 
the previously-mentioned errors in \PbPb~and in pp collisions and 
considering that the contributions due to cross section normalization, 
feed-down from beauty decays and, partially, acceptance/efficiency corrections
will cancel out in the ratio. An uncertainty of about 5\% introduced in the
extrapolation of the pp results from 14~TeV to 5.5~TeV by pQCD is also 
accounted for (see Ref.~\cite{thesis}). 

The effect of shadowing, clearly visible for $\hat{q}=0$ (no energy loss) 
as a suppression of $\RAA$, is limited to $\pt<6$--$7~\gev/c$.
Above this region only (possible) parton energy loss is expected to affect 
the nuclear modification factor of D mesons.

For $\hat{q}=4~\gev^2/\fm$ and no dead cone, we find $\RAA$ 
reduced, with respect to 1, by a factor about 3 and slightly increasing 
with $\pt$, from 0.3 at $6~\gev/c$ to 0.4 at $14~\gev/c$. Even for 
a transport coefficient lower by a factor 4, $\hat{q}=1~\gev^2/\fm$,
$\RAA$ is significantly reduced (0.5--0.6). When the dead cone effect is
taken into account, the $\RAA$ reduction due to quenching is found to
be lower by about a factor 1.5--2.5, depending on $\hat{q}$ and $\pt$. 

We point out that the estimated systematic uncertainty of about 
18\% may prevent from discriminating between a scenario with moderate 
quenching and negligible dead cone effect (e.g. $\hat{q}=1~\gev^2/\fm$ in the 
left-hand panel of Fig.~\ref{fig:RAA}) and a scenario with large quenching 
but also strong dead cone effect (e.g. $\hat{q}=4~\gev^2/\fm$ in the 
right-hand panel).

The comparison of the quenching of charm-quark-originated mesons and 
massless-parton-originated hadrons will be the best suited tool to 
disentangle the relative importance of energy loss and dead cone effects. 
The D/$charged~hadrons$ (D/$h$) ratio, defined as in (\ref{eq:RDh}), is 
presented in Fig.~\ref{fig:RDh} for the range $5<\pt<14~\gev/c$. 
We used $\RAA^h$ calculated as previously described
and $\RAA^{\rm D^0}$, 
without and with dead cone, as reported in Fig.~\ref{fig:RAA}. 
Being essentially a double ratio $\PbPb/\PbPb\times {\rm pp}/{\rm pp}$,
this observable is particularly sensitive, as many systematic uncertainties
cancel out (centrality selection and, partially, acceptance/efficiency 
corrections and energy extrapolation by pQCD). The residual systematic 
error is estimated to be of about 10--11\%.

We find that, if the dead cone correction for c quarks is not included, 
$\RDh$ is essentially 1 in the considered $\pt$ range, independently of the 
value of the transport coefficient, i.e. of the magnitude of the energy 
loss effect.
When the dead cone is taken into account, $\RDh$ is enhanced of a factor
strongly dependent on the transport coefficient of the medium:
e.g. 2--2.5 for $\hat{q}=4~\gev^2/\fm$ and 1.5 for $\hat{q}=1~\gev^2/\fm$.
The enhancement is decreasing with $\pt$, as expected (the c quark mass
becomes negligible).

The $\RDh$ ratio is, therefore, found to be enhanced, with respect to 1,
only by the dead cone and, consequently, it appears as a very 
clean tool to investigate and quantify this effect.

Since hadrons come mainly from gluons while D mesons come from (c) quarks, 
the D/$h$ ratio should, in principle, be enhanced also in absence of dead 
cone effect, as a consequence of the larger energy loss of gluons with 
respect to quarks. 
Such enhancement is essentially not observed in the obtained $\RDh$ 
because it is `compensated' by the harder fragmentation of charm quarks with 
respect to light quarks and, particularly, gluons. With $z$ the typical 
momentum fraction taken by the hadron in the fragmentation,
$\pt^{\rm hadron}=z\,\pt^{\rm parton}$, and $\Delta E$ the average
energy loss for the parton, $(\pt^{\rm parton})'=\pt^{\rm parton}-\Delta E$,
we have
\begin{equation}
  (\pt^{\rm hadron})'=\pt^{\rm hadron} - z\,\Delta E,
\end{equation}
meaning that the energy loss observed in the nuclear modification factor is, 
indeed, $z\,\Delta E$. We have, thus, to compare 
$z_{\rm c\to D}\,\Delta E_{\rm c}$
to $z_{\rm gluon \to hadron}\,\Delta E_{\rm gluon}$. With 
$z_{\rm gluon \to hadron}\approx 0.4$, $z_{\rm c\to D}\approx 0.8$ 
for $\pt^{{\rm D},h}>5~\gev/c$ 
and $\Delta E_{\rm c}=\Delta E_{\rm gluon}/2.25$ (without dead cone),
we obtain 
\begin{equation}
  z_{\rm c\to D}\,\Delta E_{\rm c}\approx 0.9\,z_{\rm gluon\to hadron}\,\Delta E_{\rm gluon}.
\end{equation} 
This simple estimate confirms that the quenching for D mesons is 
almost the same as for (non-charm) hadrons, if the dead cone 
effect is not considered. 

The errors reported in Fig.~\ref{fig:RDh} show that ALICE is expected to have
good capabilities for the study of $\RDh$: in the range $5<\pt<10~\gev/c$
the enhancement due to the dead cone is an effect of more than $3~\sigma$
for $\hat{q}>1~\gev^2/\fm$. The comparison of the values for the 
transport coefficient extracted from the nuclear modification factor of 
charged hadrons and, {\sl independently}, from the 
D/$charged~hadrons$ ratio shall provide an important test for the coherence of 
our understanding of the energy loss of hard probes propagating in the 
dense QCD medium formed in \PbPb~collisions at the LHC.

\subsection*{Acknowledgements}

I am grateful to F.~Antinori, A.~Morsch, G.~Paic,
K.~$\check{\mathrm{S}}$afa$\check{\mathrm{r}}$\'\i k, 
C.A.~Salgado and U.A.~Wiedemann 
for many fruitful discussions
and to Prof.~A.~Zichichi and Prof.~G.~'t Hooft 
for providing me with the 
opportunity to present my work in the New Talents session of the 
$41^{\rm st}$ International School of Subnuclear Physics in Erice.


\end{document}